\documentclass{svproc}
\usepackage[
pdfstartview=FitH,
CJKbookmarks=true,
bookmarksnumbered=true,
bookmarksopen=true,
colorlinks, 
pdfborder=001,   
linkcolor=black,
anchorcolor=black,
citecolor=black,
urlcolor=black
]{hyperref} 
\usepackage{moreverb,url}
\usepackage{mathrsfs}
\usepackage{amsfonts}
\usepackage{amsmath,bm}
\usepackage{amssymb}
\usepackage{graphicx}
\usepackage{epstopdf}
\usepackage{float}
\usepackage{verbatim}
\usepackage{algorithm}
\usepackage{algorithmic}
\usepackage{verbatim}

\begin{document}
	\mainmatter              
	\title{Implementation of UAV Coordination Based on a Hierarchical Multi-UAV Simulation Platform}
	\titlerunning{Multi-UAV Simulation Platform}  
	%
	\author{Kun Xiao\inst{1} \and Lan Ma\inst{2} \and Shaochang Tan\inst{3} \and Yirui Cong\inst{2} \and Xiangke Wang\inst{2}}
	\authorrunning{Kun Xiao et al.} 
	%
	\tocauthor{Kun Xiao, Lan Ma, Shaochang Tan, Yirui Cong, Xiangke Wang}
	\institute{Beijing Institute of Aerospace Systems Engineering, Beijing 100076, China,\and College of Intelligence Science and Technology, National University of Defense Technology, Changsha 410073, China \and  Flying College, Beihang University, Beijing 100191, China\\ \email{xkwang@nudt.edu.cn}}

	\maketitle              
	
	\begin{abstract}
		In this paper, a hierarchical multiple unmanned aerial vehicle(UAV) simulation platform, called XTDrone, is designed for UAV swarms, which is completely open-source\footnote{Source code at \url{https://gitee.com/robin_shaun/XTDrone}\\or \url {https://github.com/robin-shaun/XTDrone}, \\and user manual at \url{https://www.yuque.com/xtdrone/manual_en}.}. There are six layers in XTDrone: communication, simulator, low-level control, high-level control, coordination, and human interaction layers. XTDrone has three advantages. Firstly, the simulation speed can be adjusted to match the computer performance, based on the lockstep mode. Thus, the simulations can be conducted on a work station or on a personal laptop, for different purposes. Secondly, a simplified simulator is also developed which enables quick algorithm designing so that the approximated behavior of UAV swarms can be observed in advance. Thirdly, XTDrone is based on Robot Operating System(ROS), Gazebo, and PX4, and hence the codes in simulations can be easily transplanted to embedded systems. Note that XTDrone can support various types of multi-UAV missions, and we provide two important demos in this paper: one is a ground-station-based multi-UAV cooperative search, and the other is a distributed UAV formation flight, including consensus-based formation control, task assignment, and obstacle avoidance.
		\keywords{multi-UAV, hierarchical, simulation platform, UAV coordination, distributed}
	\end{abstract}
	\section{Introduction}
	Unmanned aerial vehicles (UAVs) develop rapidly due to their large potential in both civilian and military uses, such as disaster rescue, reconnaissance and surveillance. As missions becomes increasingly complex, the importance of multi-UAV cooperation grows. Therefore, a large number of researchers focuses on multi-UAV cooperation or UAV swarms in recent years \cite{maza2009multi,he2018feedback,liu2019distributed,chung2018survey,madey2012applying}.
	
	Algorithm design and validation can waste time and energy without a reliable simulation platform, and thus much attention was drawn for simulation platform developing \cite{wei2013operation,meng2015ros+,sanchez2016reliable,rodriguez2015design,garcia2009multi}. However, few platforms are open source and user friendly. Aerostack\footnote{\url{https://github.com/Vision4UAV/Aerostack}} developed from \cite{sanchez2016reliable}, is the most popular multi-UAV simulation platform according to our investigation. However, compared to single UAV simulation platforms, such GAAS\footnote{\url{https://github.com/generalized-intelligence/GAAS}} and RotorS\footnote{\url{https://github.com/ethz-asl/rotors_simulator}}, the numbers of stars and forks in github are much smaller. There is still an urgent need for a user-friendly multi-UAV simulation platform for not only researchers but also engineers. Based on this motivation, A hierarchical and modular multi-UAV simulation platform called XTDrone is developed.
	
	Considering usability, development efficiency and open source community, ROS\footnote{\url{https://www.ros.org/}}, Gazebo\footnote{\url{http://gazebosim.org/}}, PX4\footnote{\url{https://px4.io/}} and QGroundControl\footnote{\url{http://qgroundcontrol.com/}} are chosen as four bases of XTDrone. Python is the main development language and some outstanding C++ open source projects are also integrated. The platform is divided into six layers: communication, simulator, low-level control, high-level control, coordination and human interaction layers. In each layer, there are different kinds of modules, such as task allocation, Simultaneous localization and mapping (SLAM), object detection, trajectory generation, position controller and so on. All of these modules can be replaced conveniently because the input and output messages are well-defined. Therefore, platform developers can realize and test their own algorithm conveniently. Because ROS and PX4 is originally designed for embedding system, developers can deploy their algorithm to real UAVs conveniently after testing and debugging on the simulation platform. 
	
	High computational consumption is an important problem for multi-UAV simulation. Powerful workstations are not only expensive but also inconvenient compared to laptops. XTDrone runs in the lockstep mode, meaning that different numerical solvers maintain synchronized time, which makes it possible to run the simulation faster or slower than real time, and also to pause it in order to step through code. A powerful computer runs faster and less powerful one runs slower. Therefore, this feature makes computers with different performance possible to be used for simulation. Furthermore, a simplified simulator is provided, so developers can firstly test and debug their algorithm on the simplified simulator with a large-scale swarm. And then when simulating in Gazebo, they can choose to use a powerful workstation for a large-scale swarm, or reduce the number of UAVs.
		
	XTDrone is firstly a single UAV simulation platform and then a multi-UAV one. Reference \cite{xiao2020xtdrone} has provided details about single UAV simulation, and this paper focuses on the multi-UAV simulation. The rest of the paper is structured as follows. Section \ref{sec:architecture} presents the architecture of the simulation platform. And then, two demos are shown to demonstrate how the platform works. Section \ref{sec:cosearch} presents a multi-UAV cooperative search mission, planned by the ground control station. Section \ref{sec:formation_design} and Section \ref{sec:formation_simulation} presents a distributed UAV formation, including consensus cooperative control, task assignment and obstacle avoidance. Section \ref{sec:formation_design} describes algorithm designs and Section \ref{sec:formation_simulation} presents its simulation implementation. Section \ref{sec:conclusion} concludes the paper and indicates future work.

	\section{Architecture}
	\label{sec:architecture}
\begin{figure}[H]
		\begin{center}
			\includegraphics[
			width=12cm]{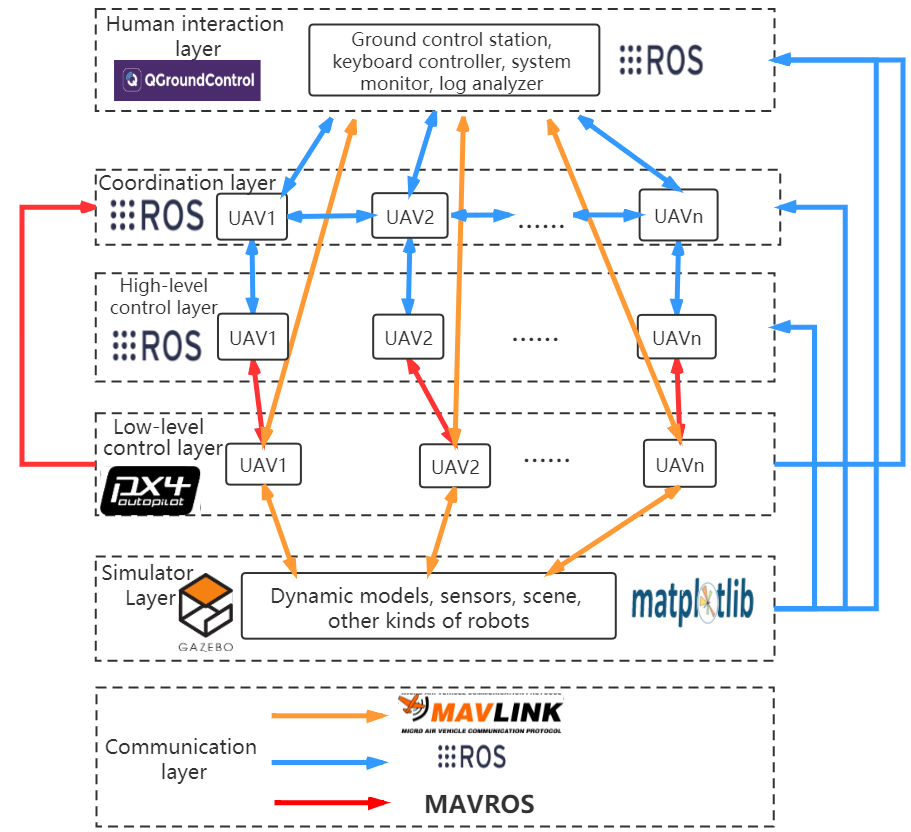}
		\end{center}
		\caption{Architecture of XTDrone}
		\label{fig:architecture}
	\end{figure}

	Fig. \ref{fig:architecture} shows the architecture of XTDrone. Five layers communicate with each other through the communication layer. The architecture is inspired by \cite{liu2019mission}, which is a a multi-layered and distributed
	architecture for real UAV swarm. To some extent, the architecture of XTDrone is a simulation version of that in \cite{liu2019mission}. In this section, the six layers are introduced respectively from down to top. 
	
	The communication layer is in charge of the messages transmission among all the UAVs and the ground control station, which is the base of all other layers. ROS, MAVLink\footnote{\url{https://mavlink.io/en/}} and MAVROS\footnote{\url{https://github.com/mavlink/mavros}}, supply message-passing in the communication layer. To be user-friendly, XTDrone encapsulates complex protocols, and standardizes the topic names. Developers can conveniently subscribe and publish the needed topics in their designed modules.  
		
	The simulator layer supplies UAV dynamic models, sensors, scene and other kinds of robots, all of which is customizable, so that developers can modify the aerodynamics model, or add multiple cameras to the UAV according to their needs. The main simulator is Gazebo. Moreover, for quick algorithm developing in the early stage, a simplified simulator based on Matplotlib\footnote{\url{https://matplotlib.org/}} is provided. Reference {\cite{xiao2020xtdrone}} presents more details about this layer. 
	
	Low-level control layer is totally based on PX4 software in the loop (SITL), which contains state estimation and low-level control such as position control, attitude control and flight modes. XTDrone focuses more on high-level control layer and coordination layer, and developers can just use this layer without extra modification. To develop algorithm in this layer, developers are expected to be familiar with PX4.  
	
	High-level control layer contains perception and motion planning. It's a key layer for intelligence of a single UAV. Object detection and tracking, SLAM, obstacle avoidance, etc. are all in this layer. Some demos have been realized in XTDrone, and developers can modify them or rewrite them according to their needs. For a UAV swarm, this layer receives tasks allocated from coordination layer, and then each UAV achieves the task.
	
	The coordination layer is responsible for the tasks related to negotiation (e.g., task allocation) among UAVs for mission coordination. It divides the total mission into different small missions, and then sends them to high-level control layer. This layer is very flexible, and may contain different modules according to different missions and coordination strategies. For example, in the formation demo (Section \ref{sec:formation_design} and \ref{sec:formation_simulation}), the task assignment and the consensus controller is in this layer. 
	
\begin{figure}[H]
		\begin{center}
			\includegraphics[
			width=12cm]{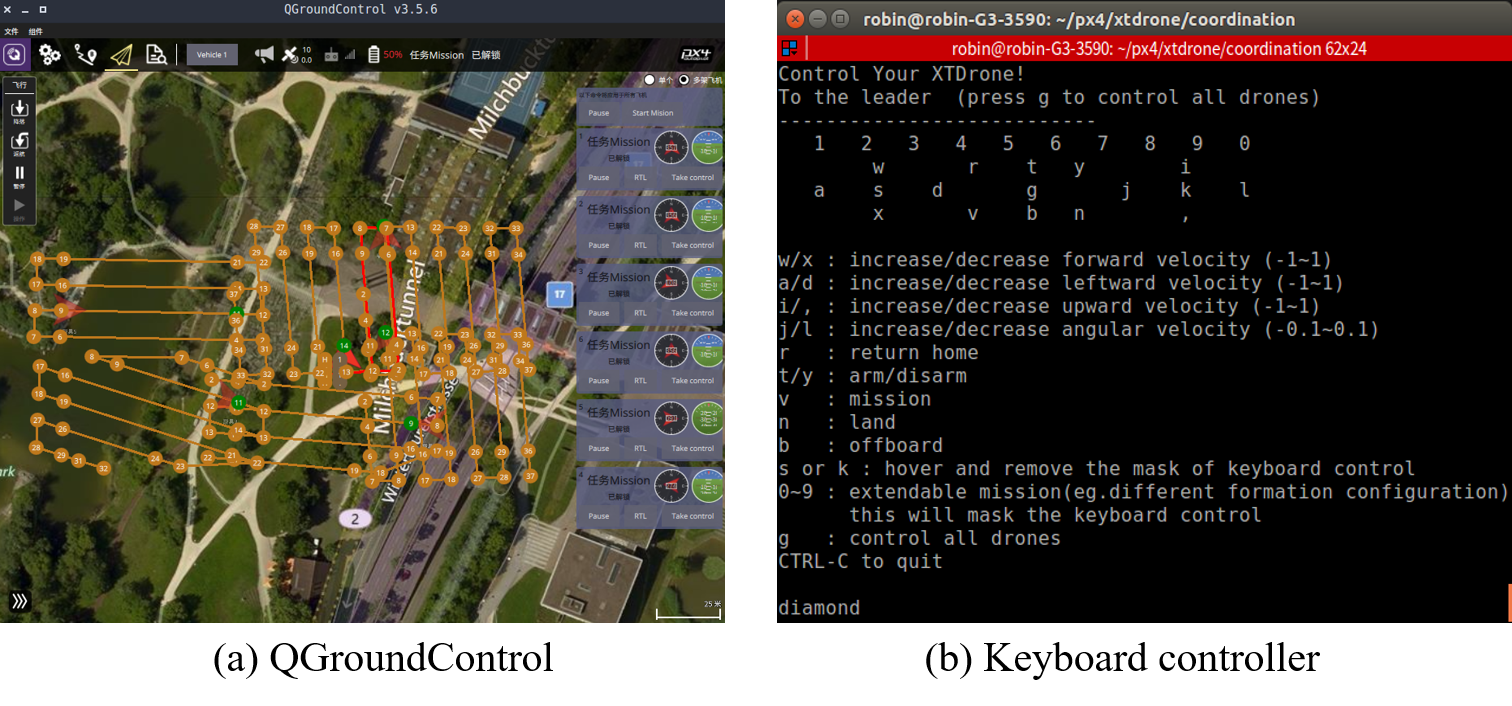}
		\end{center}
		\caption{Ground control station and keyboard controller}
		\label{fig:layer6}
	\end{figure}

	The human interaction layer is a set of interfaces for the developer to control and monitor the UAV swarm. For UAVs, a typical human machine interface is  ground control station, and specifically, XTDrone utilizes QGroundControl, shown as Fig. \ref{fig:layer6}(a). QGroundControl communicates with PX4 SITL through MAVLink, so developers can monitor and adjust parameters and targets of low-level control layer. Besides the low-level layer, coordination layer and high-level layer are controlled and monitored through a set of tools based on ROS, such as a customizable keyboard controller, shown as Fig. \ref{fig:layer6}(b),  rviz\footnote{\url{https://github.com/ros-visualization/rviz}} and rqt\footnote{\url{https://github.com/ros-visualization/rqt}}. Moreover, logging is essential for analyzing the algorithm and debugging the code. For ROS, ROSBag is a useful tool to record logs, and for PX4, flight logs, containing almost every key message, are recorded as ULog files\footnote{\url{https://dev.px4.io/master/en/log/ulog_file_format.html}}. And there are some user-friendly software\footnote{\url{https://docs.px4.io/master/en/log/flight_log_analysis.html}} to analysis ULog files.
	
	\section{Cooperative Search Simulation}
	\label{sec:cosearch}

	In this section, a multi-UAV cooperative search is implemented. We plan the trajectories for six UAVs through QGroundControl, the low-level controller of each UAV tracks the desired trajectory and the high-level controller of each UAV detects the target objects.
	
\begin{figure}[H]
		\begin{center}
			\includegraphics[
			width=12cm]{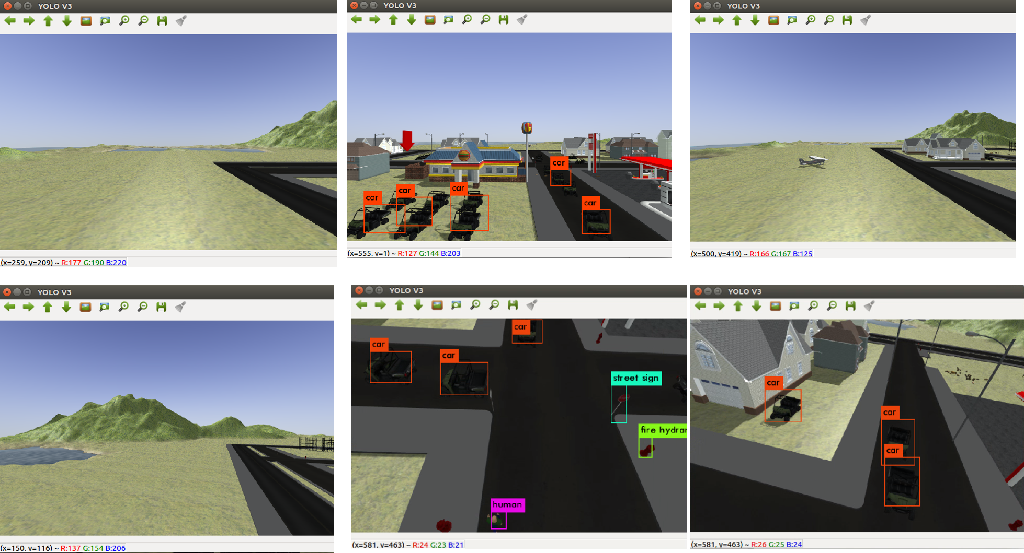}
		\end{center}
		\caption{Human searched by six UAVs }
		\label{fig:yolo}
	\end{figure}

	The mission goal is to search a human in a large region. A tiny-YoloV3 network \cite{redmon2018yolov3} is trained beforehand, and deployed in the high-level controller layer\footnote{\url{https://www.yuque.com/xtdrone/manual_en/target_detection_tracking}}. For searching in a wide region, a simple but efficient way is to divide the region into six non-overlapping regions. Each UAV searches one region and detects the target objects. By setting and uploading way points in QGroundControl, shown as Fig. \ref{fig:layer6}(a), the low-level layer of each UAV tracks the desired flight trajectory, and meanwhile, tiny-YoloV3 works, shown as \ref{fig:yolo}. Finally, the NO.5 UAV detects the human and then the mission is completed.
	
	
	\section{Distributed UAV Formation Design}
	\label{sec:formation_design}

	In this section, a design of distributed UAV formation control is represented. A consensus controller, a task assignment strategy and an obstacle avoidance algorithm are designed. The formation is a leader-following formation. Firstly, the leader assigns task to all the followers. And then for each follower, the output velocities of the consensus controller and the obstacle avoidance module are composed to desired velocity in the high-level control layer. And then the desired velocity is sent to the low-level control layer. Finally, the desired formation is achieved.
	
	\subsection{Consensus-based Formation}
	Considering a group of $N$ UAVs, each can be modeled as an first-order integrator model on the basis of its kinematic model, then it can be described as:
	\begin{equation}
	\dot{\xi_i}=u_i,  \qquad i=1,2,\dots ,N \label{basic_model}
	\end{equation}
	where the input $u_i\in \mathbb{R}^3$ is the velocity of UAV $i$. $\xi_i\in \mathbb{R}^3$ is the position of UAV $i$\. The target state is that all the UAVs achieve a given formation pattern. The formation controller is given as follows:
	\begin{equation}
	u_i=-\sum_{j=1}^{N} w_{ij} \{[(\xi_i - \delta_i )- (\xi_j - \delta_j) ]\} \label{consensus_model}
	\end{equation}
	where $ w_{ij}$ is the  $(i,j)$ element of the adjacency matrix $ W=[w_{ij}] \in \mathbb{R}^{N\times N}$, which represents that UAV $j$ communicates with UAV $i$ if and only if $ w_{ij} >0$. $\delta_i$ denotes the formation offset of UAV $i$ and it is determined by the desired formation configuration.
	
	We consider that system \ref{consensus_model} achieves consensus when $\parallel \xi_i - \delta_i )- (\xi_j - \delta_j)\parallel \to 0 $. Because the consensus of the system and the stability of the formation are equivalent, our system could achieve formation stability in this condition \cite{ren2008distributed}.

\begin{figure}[H]
		\begin{minipage}[t]{0.4\linewidth}
			\begin{center}
				\includegraphics[
				width=2cm]{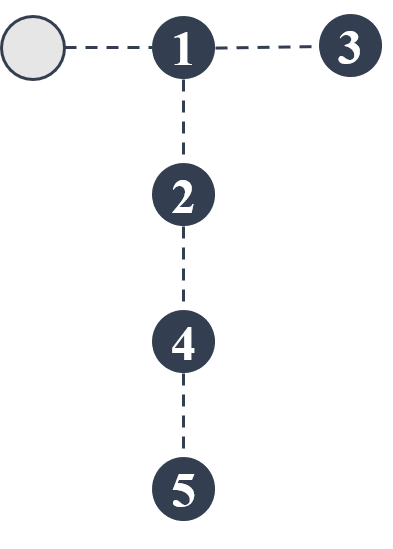}
			\end{center}
			\caption{Current position of 6 UAVs}
			\label{uav_6_T_position}
		\end{minipage}
		\hfill
		\begin{minipage}[t]{0.6\linewidth}
			\begin{center}
				\includegraphics[
				width=7cm]{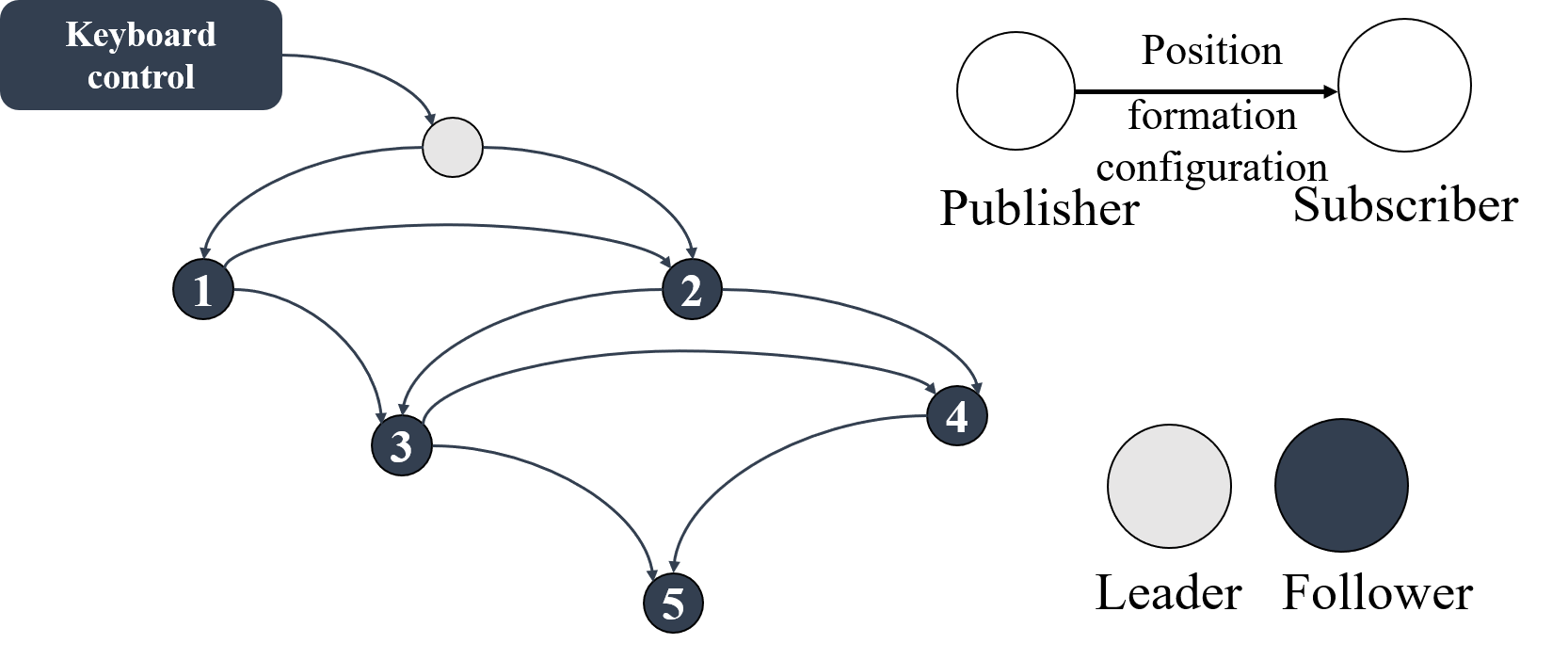}
			\end{center}
			\caption{Communication topology of 6 UAVs}
			\label{uav_6_topology}
		\end{minipage}
	\end{figure}

	In \cite{8822694}, Yu Ding built an interaction topology based on their proposed ``Veteran Rule". We use this to determine the communication topology of UAVs, e.g. the adjacency matrix $W=[w_{ij}]$. Assume that the number of UAVs is 6, and the current position is ``T" shown in Fig.\ref{uav_6_T_position}, then we get the communication topology as Fig.\ref{uav_6_topology} and the adjacency matrix $\mathcal{L}$ as \eqref{L_uav_6}
	
	\begin{equation}
	\mathcal{L}=\begin{bmatrix}
	0 & 0 & 0 & 0 & 0 & 0 \\
	1 & 0 & 0 & 0 & 0 & 0 \\
	1 & 1 & 0 & 0 & 0 & 0 \\
	0 & 1 & 1 & 0 & 0 & 0 \\
	0 & 0 & 1 & 1 & 0 & 0 \\
	0 & 0 & 0 & 1 & 1 & 0 \\
	\end{bmatrix}
	\label{L_uav_6}
	\end{equation}
	
	\subsection{Task assignment}
\begin{figure}[H]
		\begin{center}
			\includegraphics[
			scale=0.35]{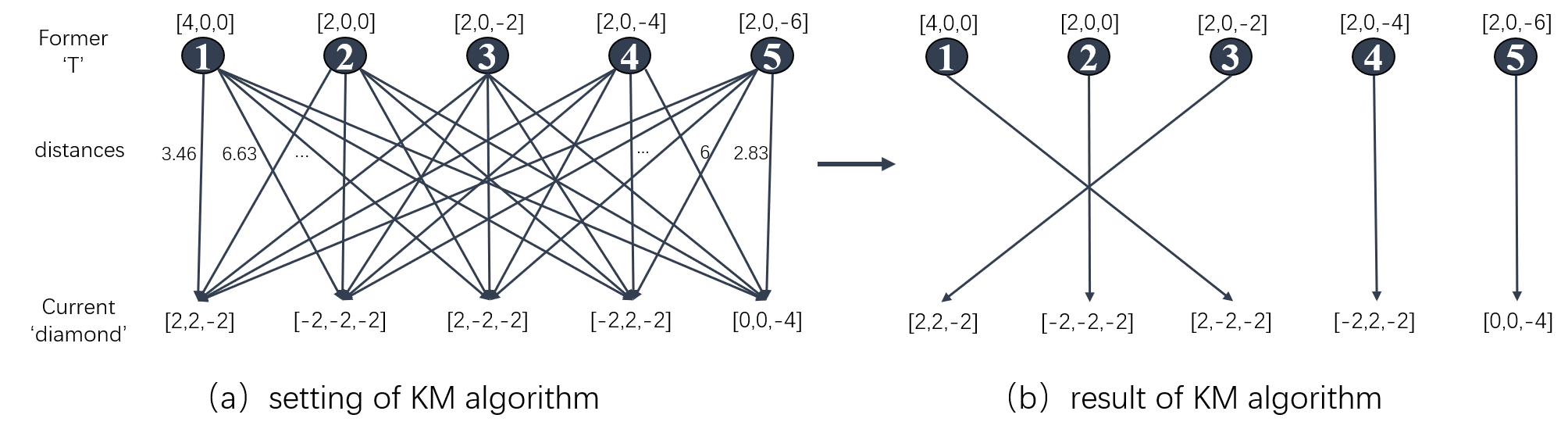}
		\end{center}
		\caption{Settings and results of KM algorithm}
		\label{KM_assignment}
	\end{figure}

	We consider formation reconfiguration as a kind of task assignment which setts different UAVs into different formation positions for the shortest total distance while changing formation. We adopt Kuhn-Munkres (K-M) algorithm for task assignment. As the current formation offset and the target formation offset could be obtained by each UAV, we set them as a bipartite graph. Each weight of graph is the distance (in practice turned to negative ) between the current formation offset and the target formation offset. The assignment solved by KM algorithm obtains the best total performance with the shortest total distance\cite{6084856}. For instance, for the formation reconfiguration from `T' to `diamond', the settings and results of KM algorithm are shown in Fig.\ref{KM_assignment}
	
	\subsection{Obstacle avoidance}	
	
\begin{figure}[H]
		\begin{center}
			\includegraphics[
			width=7cm]{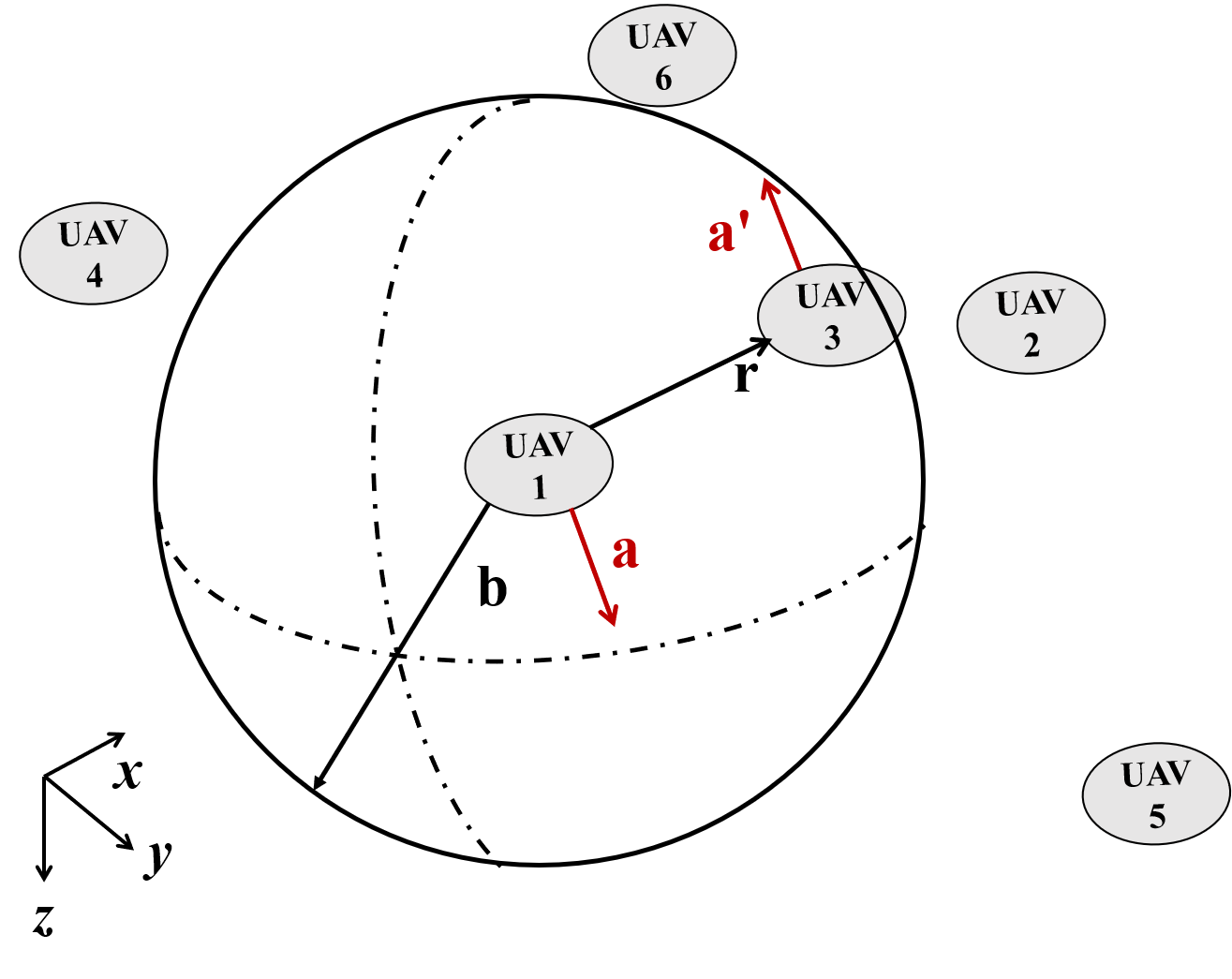}
		\end{center}
		\caption{Strategy of obstacle avoidance}
		\label{fig:obstacle_avoidance}
	\end{figure}

	Fig. \ref{fig:obstacle_avoidance} shows the strategy of obstacle avoidance. A UAV will avoid when another UAV is in the range of $b$ meters. $\mathbf{r}$ is the distance vector and points to the obstacle UAV. $\mathbf{a}$ is the avoidance vector of UAV1 and  $\mathbf{a'}$ is the avoidance vector of UAV3.  $\mathbf{a}$ and  $\mathbf{a'}$ are in the opposite direction, and perpendicular to $\mathbf{r}$. Because there are infinite vectors perpendicular to $\mathbf{r}$, a simple method, shown as Algorithm \ref{alg:obstacle_avoidance} is designed to identify a unique solution. Two auxiliary vectors $\mathbf{n}_1$ and $\mathbf{n}_2$ are used to avoid the cross product result close to zero vector. $kp$ is a proportion factor. If there are more than one UAVs in the range of $b$ meters, avoidance vector $\mathbf{a}$ will be modified several times. The final $\mathbf{a}$ will be added to the desired velocity, and then the desired velocity will be sent to the low-level control layer.
	
	Another question is how to obtain the relative position of other UAVs. Because of the distributed communication, a UAV cannot obtain all other UAVs' absolute positions. There are many researches focused on the cooperative relative localization algorithm \cite{xu2020decentralized,saska2017system,guo2017ultra}. In this demo, for simplification, the relative position ground truth is sent to each UAV. Developers can just add a relative localization module to replace the ground truth with the module output.  
	\begin{algorithm}
		\begin{footnotesize}
			\caption{Obstacle avoidance}\label{alg:obstacle_avoidance}
			\begin{algorithmic}[1]
				\LOOP
				\STATE  \textbf{Initialization:} $\mathbf{a}_0 = [0, 0, 0]$; $\mathbf{n}_1 = [1, 0, 0]$; $\mathbf{n}_2 = [0, 1, 0]$; $i = 1$ 
				\WHILE {$i \leq \text{number of UAVs in the range of}$ $b$ $\text{meters}$}
				\IF {$\mathbf{r}_i\cdot{\mathbf{n}_1} < \mathbf{r}_i\cdot\mathbf{n}_2$}
				\STATE $\mathbf{a}_0 = \mathbf{a}_0 + kp * (1 - |\mathbf{r}_i|/b) * (\mathbf{r}_i\times{\mathbf{n}_1} / |\mathbf{r}_i\times{\mathbf{n}_1}|)$;
				\ELSE
				\STATE $\mathbf{a}_0 = \mathbf{a}_0 + kp * (1 - |\mathbf{r}_i|/b) * (\mathbf{r}_i\times{\mathbf{n}_2} / |\mathbf{r}_i\times{\mathbf{n}_2}|) $;
				\ENDIF
				\STATE $i = i + 1$;
				\ENDWHILE
				\ENDLOOP
			\end{algorithmic}
		\end{footnotesize}
	\end{algorithm}
	
	\section{Distributed UAV Formation Simulation}
	\label{sec:formation_simulation}
	XTDrone contains two python classes, one for the leader and another for the follower. Developers can inherit these two classes to modify the communication and control scheme. By starting multiple ROS nodes, multiple UAV controllers are simulated. 

\begin{figure}[H]
		\begin{center}
			\includegraphics[
			width=12cm]{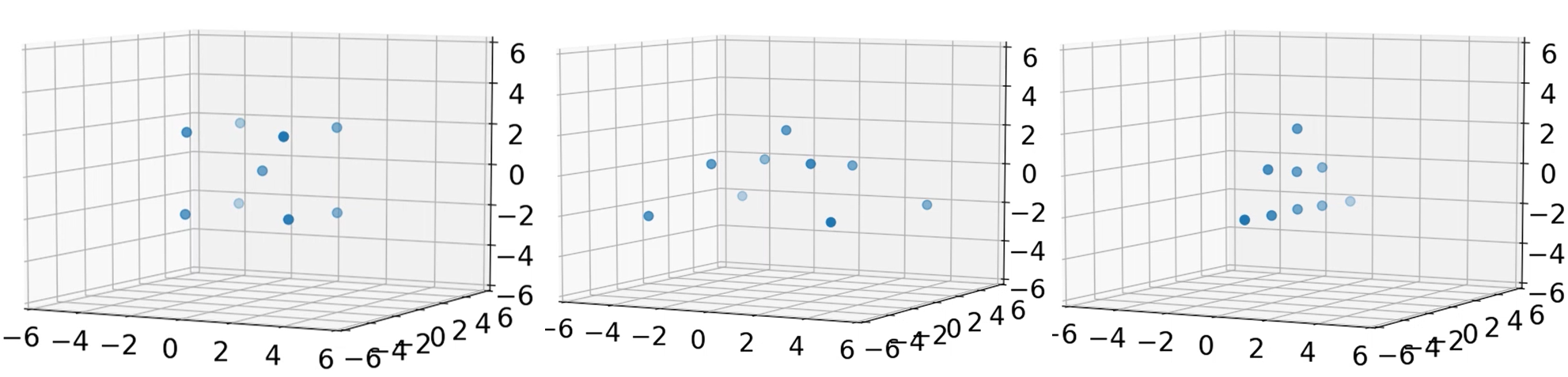}
		\end{center}
		\caption{Formation of 9 UAVs implemented in the simplified simulator}
		\label{simple_simulator}
	\end{figure}

\begin{figure}[H]
		\begin{center}
			\includegraphics[
			width=12cm]{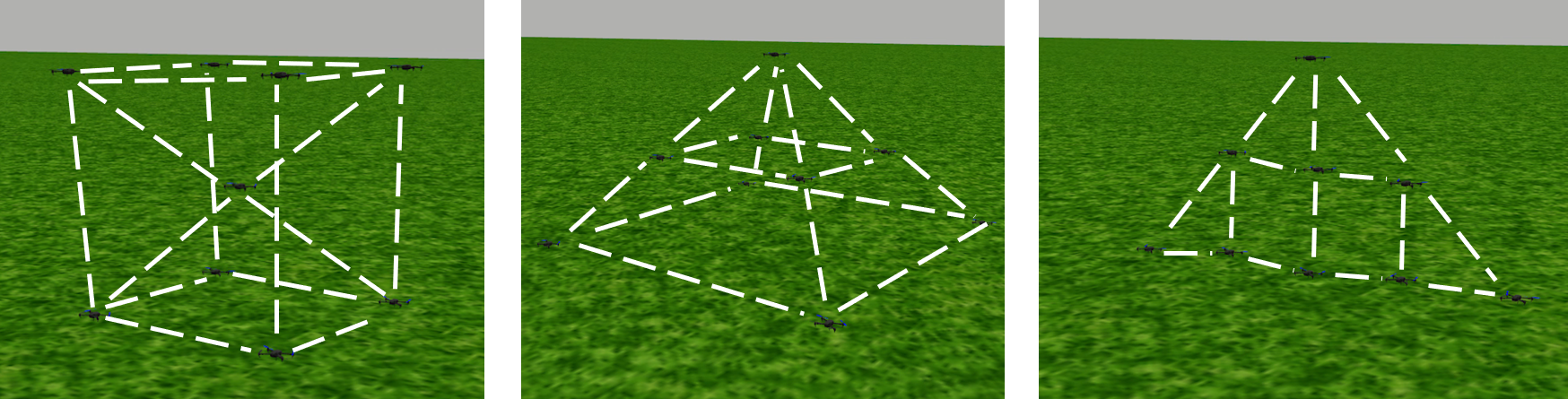}
		\end{center}
		\caption{Formation of 9 UAVs implemented in Gazebo}
		\label{gazebo_total_changing}
	\end{figure}

	Fig. \ref{simple_simulator} shows the formation of 9 UAVs implemented in the simplified simulator and  Fig. \ref{gazebo_total_changing} shows the formation implemented in Gazebo. Three configurations from left to right are cube, pyramid and triangle. The videos of the formation reconfiguration can be seen at \url{https://www.yuque.com/xtdrone/demo/uav_formation}.
	
\begin{figure}[H]
		\begin{center}
			\includegraphics[
			width=12cm]{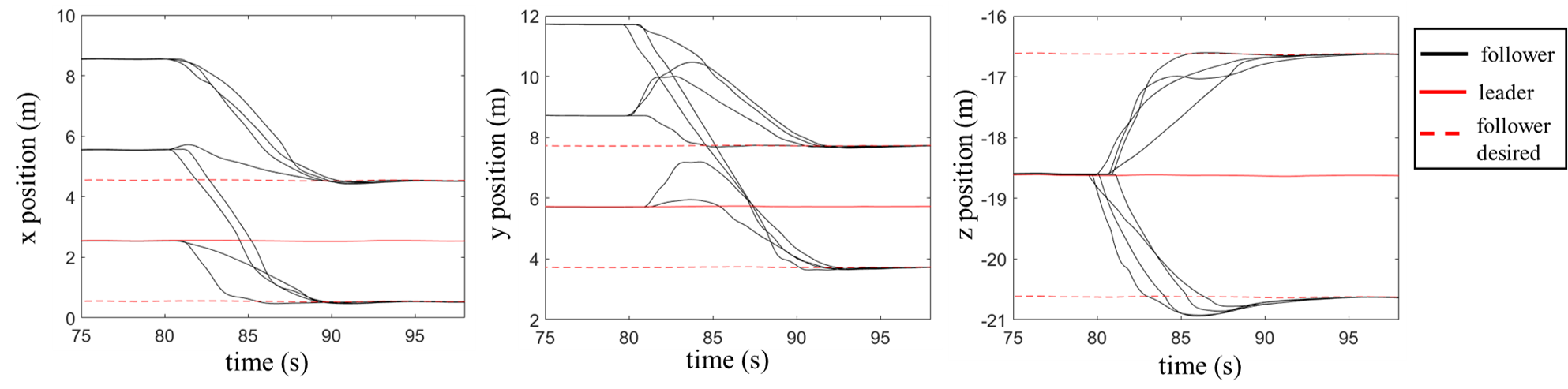}
		\end{center}
		\caption{Position response curves}
		\label{fig:position_response}
	\end{figure}

	To validate the consensus control algorithm,  a flight log is recorded during configuration from origin to `cube', and the position response curves are shown in Fig. \ref{fig:position_response}. Although some overshooting can be seen on the response curves, all the UAVs reaches stability.  

	\section{Conclusion and Future Work}
	\label{sec:conclusion}
	A hierarchical multi-UAV simulation platform is developed. Six layers: communication, simulator, low-level control, high-level control, coordination and human interaction layers are designed. Two demos are presented to demonstrate how the platform works. The first one is a cooperative search mission. The low-level controller of each UAV tracks the desired trajectory planned by ground control station and the high-level controller of each UAV detects the target objects. The other one is a distributed UAV formation including consensus control, task assignment and obstacle avoidance. The simulation validation contains 9 UAVs simulation in the simplified simulator and in Gazebo, which validates the consensus control algorithm.  
	
	The platform is all open source, and is being developed continually. Now it only supports multi-rotor UAVs and the support to fixed-wing UAVs and vertical take-off and landing UAVs is coming. Besides, another simulator, AirSim, is being integrated into the platform. Furthermore, a multi-UAV competition is planned to be held based on the simulation platform.
	
	\section*{Acknowledgement}
	This work was supported by the National Natural Science Foundation of China under Grants 61973309 and 61801494.
	
	%
	%
	\bibliographystyle{bibtex/spmpsci_unsrt}
	
	\bibliography{reference}
\end{document}